\documentclass{article}

\usepackage[final]{nips_2018} 

\usepackage[final]{nips_2018}

\usepackage[utf8]{inputenc} 
\usepackage[T1]{fontenc}    
\usepackage{hyperref}       
\usepackage{url}            
\usepackage{booktabs}       
\usepackage{amsfonts}       
\usepackage{nicefrac}       
\usepackage{microtype}      
\usepackage[pdftex]{graphicx}   
\usepackage{float}
\usepackage{url}
\usepackage{subfigure}
\title{ShanshuiDaDA: An Interactive, Generative System towards Chinese Shanshui Painting}

\author{
Aven Le Zhou$^{1,2}$, Qiufeng Wang$^{2}$, Cheng-Hung Lo$^{2}$, Kaizhu Huang$^{2}$\\
  $^1$artMachines $^2$Xi'an Jiaotong-Liverpool University\\
  \texttt{hi@aven.cc, \{Qiufeng.Wang, CH.Lo, Kaizhu.Huang\}@xjtlu.edu.cn} \\
  \And
}

\begin{document}

\maketitle
\begin{abstract}
“Shanshui” literally means “mountain and water”, is an East Asian traditional type of brush painting involves natural landscape. In this paper, we propose an interactive and generative system based on Generative Adversarial Network(GAN), which helps user draw "Shanshui" easily. We name this system and installation “ShanshuiDaDA”. “ShanshuiDaDA” is trained with “CycleGAN” and wrapped with a web-based interface. When participants scribble lines and sketch the landscape, the “ShanshuiDaDA” will assist them to generate/create a Chinese "Shanshui" painting in real-time.
\end{abstract}

\section{Introduction}
“Shanshui” literally means “mountain and water”, also known as literati painting, it's an East Asian type of brush painting of Chinese origin that uses ink and involves natural landscape. As a key element of what Chinese calls literati arts — or amateur arts of the scholars, “Shanshui” used to be along with their education, the Chinese scholars were all trained in this forms of fine arts. This art form is, in a long history, an essential part of the spiritual life of the entire community of ancient Chinese intellectuals [1]. But the tradition is vanishing. “DaDA” refers to “Design and Draw with AI”, explains the goal of exploring the possible role of artificial intelligence in (traditionally human-oriented) creative processes -- such as drawing and design. 

In this project, we choose “Shanshui” – this unique eastern traditional art form to train "ShanshuiDaDA" which learns a mapping from hand sketch to “Shanshui”. In addition, we deliver the project in an interactive flavor which commits to enhance modern people, especially those have grown up in eastern culture, with the ability to easily use “Shanshui” as an expressive medium and to enrich their spiritual life.

\begin{figure}[H] 
	\centering    
	\subfigure[Interface]{
		\begin{minipage}{3cm}
			\centering
			\includegraphics[width=3.4cm]{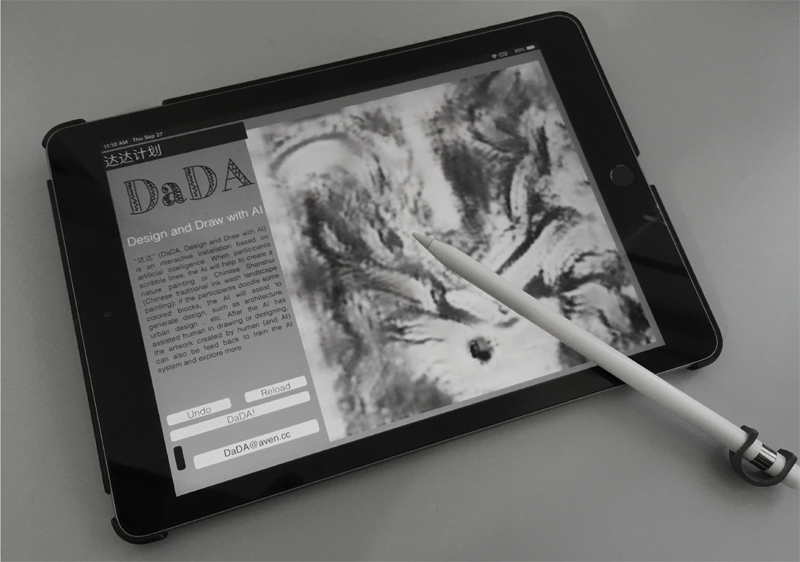}
		\end{minipage}
	}       
	\subfigure[Installation Setup]{
		\begin{minipage}{3cm}
			\centering
			\includegraphics[width=3.4cm]{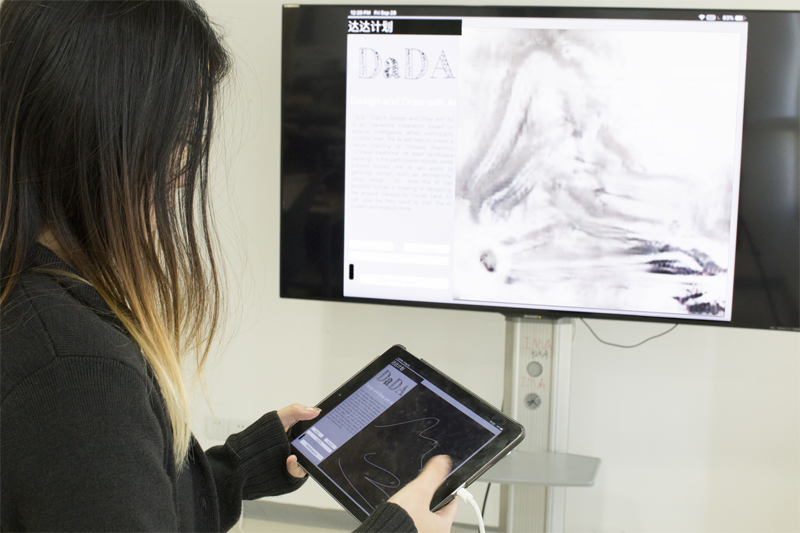}
		\end{minipage}
	}       
	\subfigure[Collection of Generated "Shanshui"]{
		\begin{minipage}{3cm}	
			\centering
			\includegraphics[width=3.4cm]{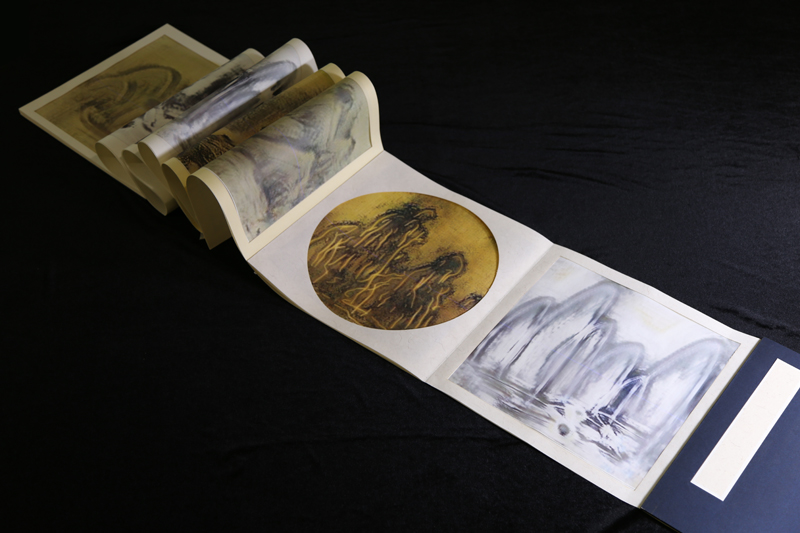}
		\end{minipage}     
	}     
	\caption{"Shanshui-DaDA"}     
	\label{figure1}     
\end{figure}

\section{System Overview}
\label{headings}
\textbf{Generative:} Machine learning tasks on Sketch‒to‒Shanshui transition are not found in the scope of this project, there are also very few available data sets. Thus, we collected and made the data set. This includes collecting “Shanshui” copies and pre-processing the scanned copies as well as creating the hand sketches. All “Shanshui” paintings used in this project are high-resolution scans available on the open data platform of the National Palace Museum [2].
\begin{figure}[H]
	\centering
	\includegraphics[width=12cm]{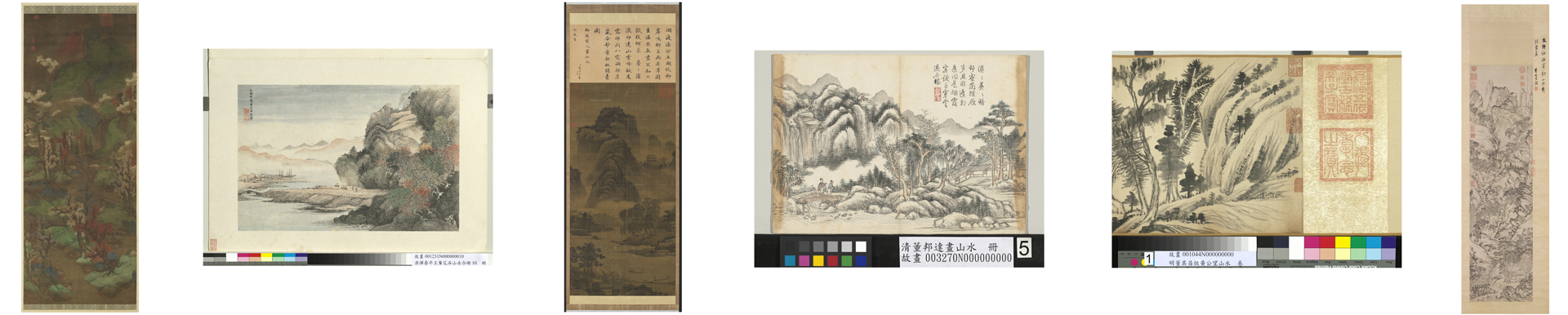}
	\caption{Sample of Collected “Shanshui” Scans}
	\label{figure2} 
\end{figure}
The sketch data set is created through computer vision methods. After cropped off the frames, we applied a canny filter to each painting and generated the edges as hand sketch data [3]. We then trained on the data set with “CycleGAN” algorithm and obtained the sketch-to-Shanshui model [4]. 
%

\textbf{Interactive:} In order to create a real‒time and interactive experience, hand sketch from the participant is fed in the pre-trained sketch-to-Shanshui model and generates immediate feedback and presents in the interface. The main architecture of this installation is a web‒based client and server system. The "p5.js" web page works as front-end interface and runs on an iPad (Figure 1.a). Participants will draw on the canvas and submit the hand-sketch to back-end server which runs on the cloud. The server will handle data communication and execute command for running the pre-trained model in a test mode. The generated painting will then be post to the front interface. In the installation setup, the front interface (the iPad) is connected to a TV screen for better visuals (Figure 1.b). 

%
%

%

\section{Result and Discussion}
\begin{figure}[H]
	\centering
	\includegraphics[width=12cm]{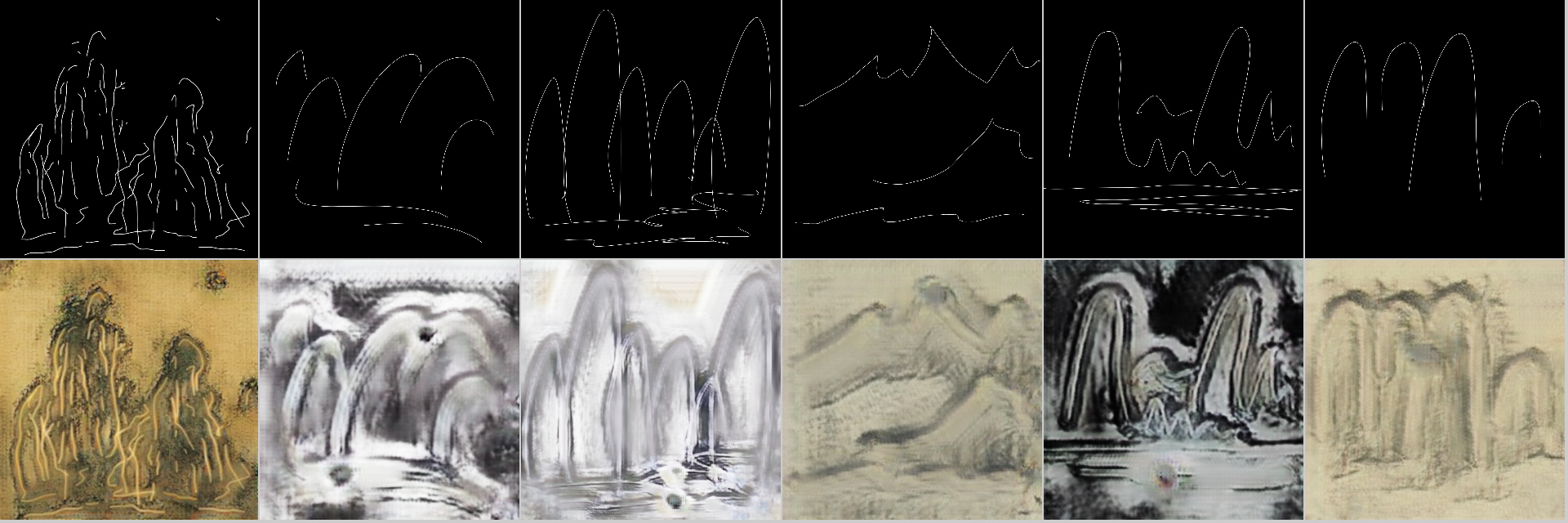}
	\caption{Selected Generated “Shanshui”}
	\label{figure3} 
\end{figure}
In figure 3, there are six selected pairs of hand sketches and corresponding generated "Shanshui" paintings from participants. The “ShanshuiDaDA” obviously learned various different styles. Some can be easily mapped to the existed ones, like the first column in figure 3 appears very like the “Qingbi Shanshui” before Tang Dynasty, but some can be hardly categorized. Such as the generated "Shanshui" in third column, it presents a strong ink-wash-painting style but matches rarely any previous styles. In other words, “ShanshuiDaDA” has even created a brand new style.

The interactive, generative progress in this project demonstrates a cooperative relationship between human and AI. Human creator not only trains AI with artificial data but also benefits from the assistance of AI. On the other side, artificial intelligence not only learns from human-created data but also "teaches" and provides human creator new approaches to creative goals.


\section*{References}
\medskip
\small

[1] W. Shi.\ Shan shui in the world: A generative approach to traditional chinese
landscape painting. In {\it VIS 2016 Arts Program}, 2016.

[2] Palace Museum at Taibei Open Data {\it $@$Online}.\ \url{https://theme.npm.edu.tw/opendata/}, 2018.

[3] Canny J. A computational approach to edge detection[J]. IEEE Transactions on pattern analysis and machine intelligence, 1986 (6): 679-698.

[4] J.-Y. Zhu, T. Park, P. Isola, and A. A. Efros.\ Unpaired image-to-image
translation using cycle-consistent adversarial networks. In {\it ICCV}, 2017.
\section*{Appendix}
We selected 24 generated paintings and printed out in a tradtional "Shanshui" book design. There are also more video demonstration and documentation can be found at: www.aven.cc/ShanshuiDaDA.

\begin{figure}[H]
	\centering
	\includegraphics[width=11.5cm]{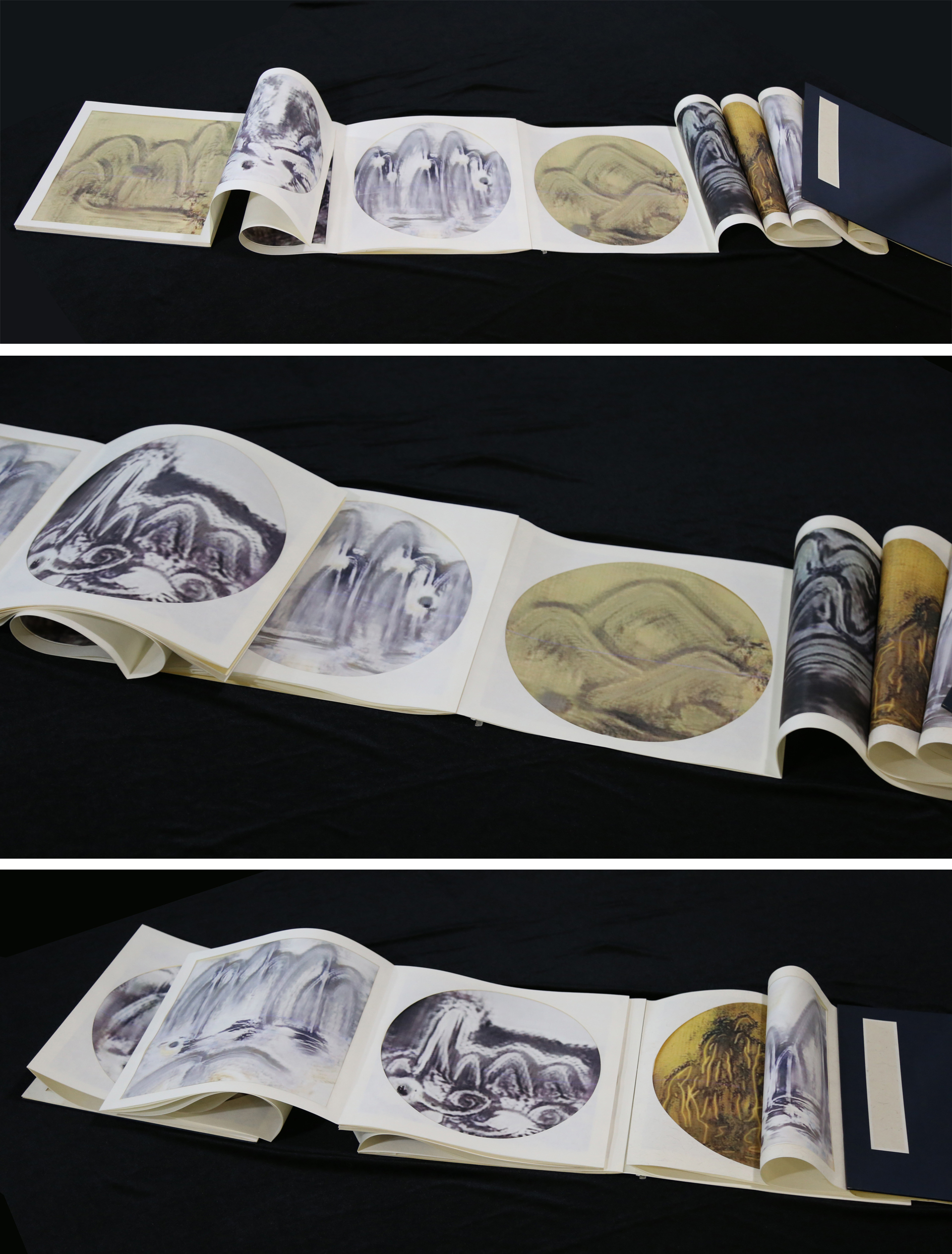}
	\caption{Selected Generated Paintings Printed in Traditional Design of "Shanshui" Collection}
	\label{figure4} 
\end{figure}

\begin{figure}[H]
	\centering
	\includegraphics[width=11cm]{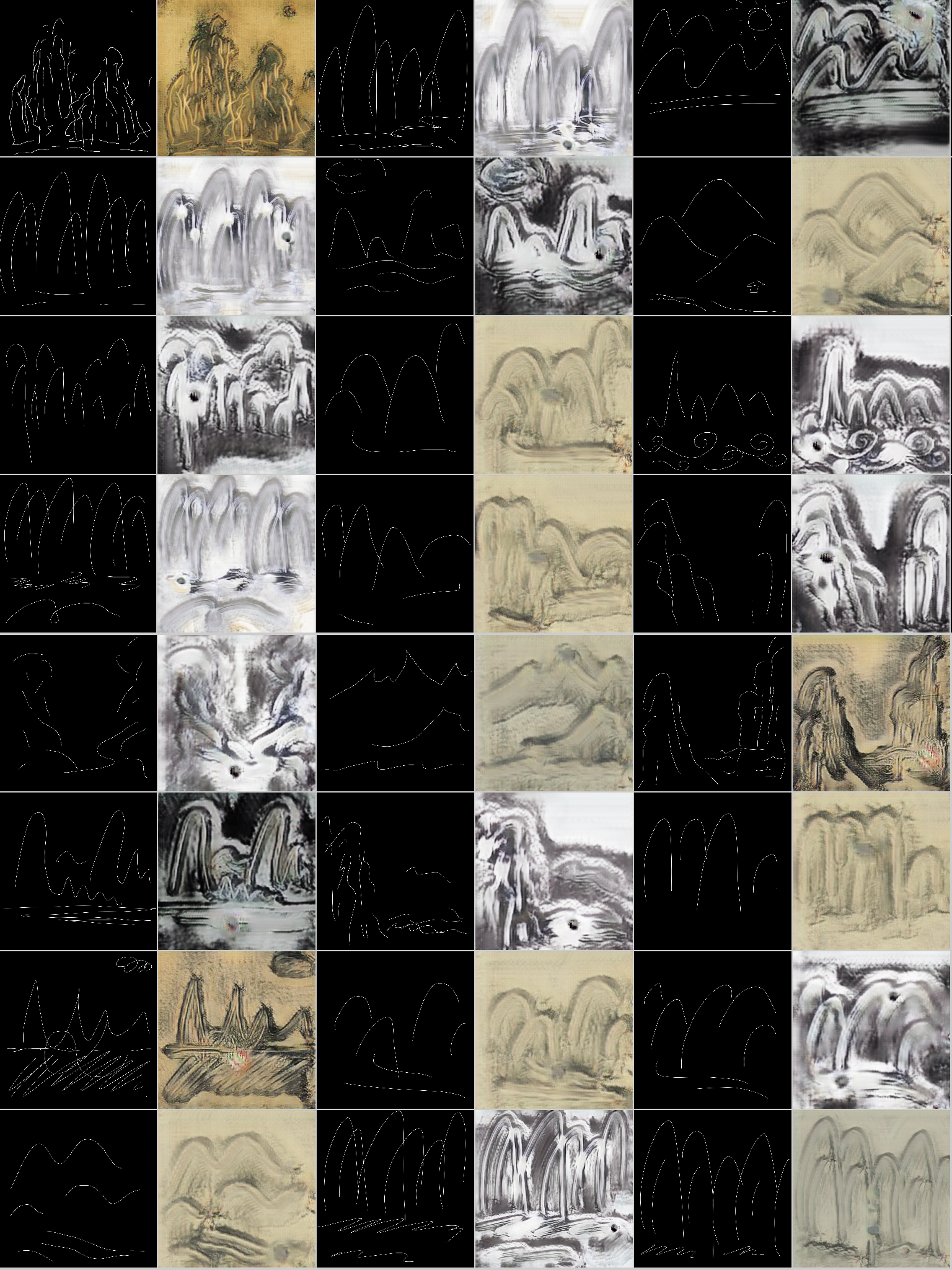}
	\caption{Generated “Shanshui” with Participants' Sketches Collection, Book Resources}
	\label{figure5} 
\end{figure}

\begin{figure}[H]
	\centering
	\includegraphics[width=11cm]{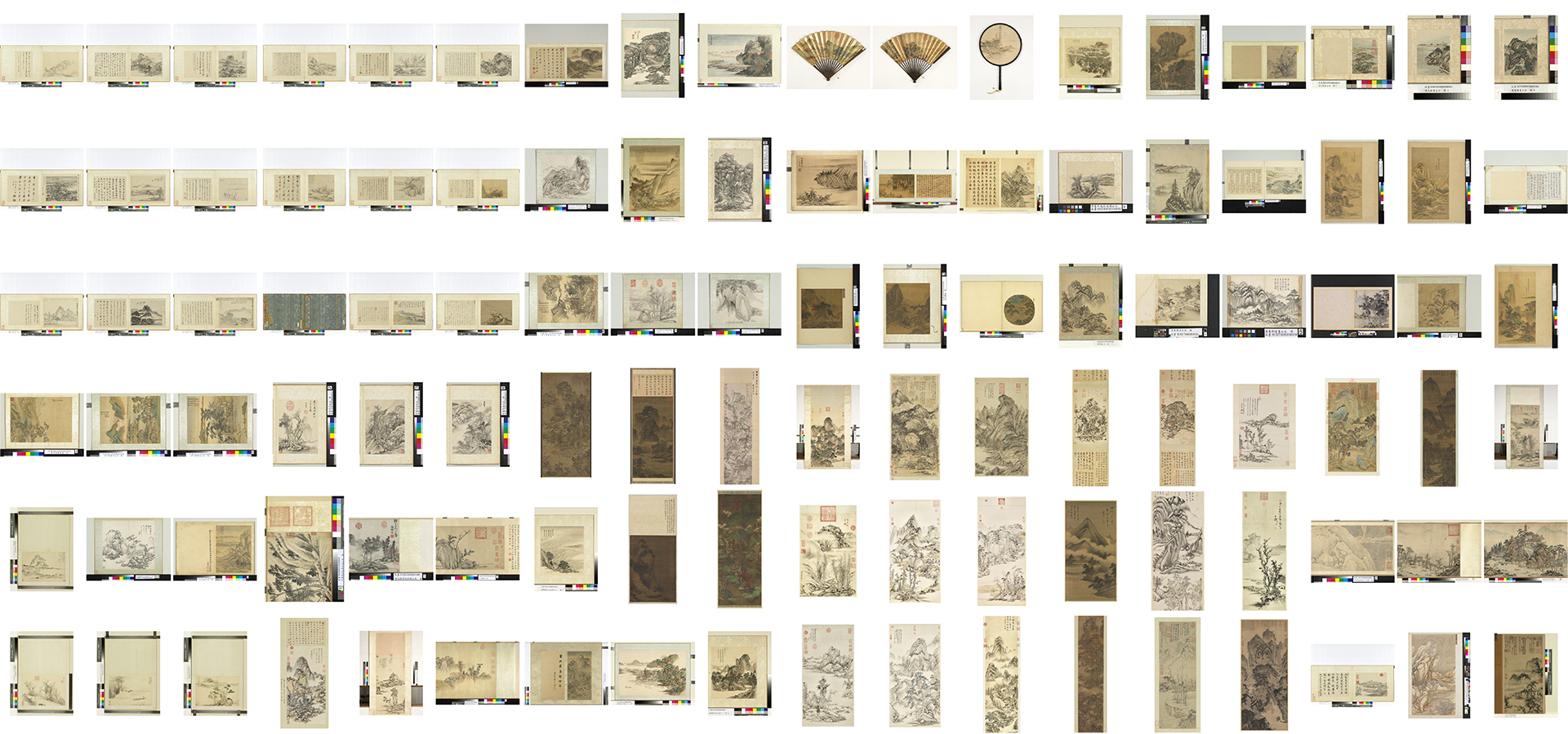}
	\caption{Collected Raw “Shanshui” Used in this Project}
	\label{figure6} 
\end{figure}
\end{document}